\documentclass[oneside,11pt,a4paper]{article}
\pdfoutput=1
\usepackage{jheppub}
\usepackage[utf8]{inputenc}
\usepackage[T1]{fontenc}

\usepackage{etex}
\usepackage{amsmath}
\usepackage{amssymb}
\usepackage{lmodern}
\usepackage{graphicx}
\usepackage{microtype}

\usepackage[]{caption}
\usepackage{booktabs}

\usepackage{multicol}
\usepackage{longtable}
\usepackage{float} 
\usepackage{multirow}

\usepackage{xcolor}
\definecolor{darkgreen}{rgb}{0.0, 0.6, 0.2}

\usepackage{paralist}
\usepackage{bbm}
\usepackage{tikz}
\usetikzlibrary{patterns}

\usepackage{dsfont}
\usepackage{braket}
\usepackage{units}
\usepackage{mathtools}
\usepackage{nicefrac}
\usepackage{slashed}

\advance \headheight by 3.0truept       

\usepackage{fancyhdr}
\newcommand{\I}{\mathrm{i}}
\newcommand{\Mod}{\ensuremath{\;\mathrm{mod}\;}}

\newcommand{\Id}{\mathbbm{1}}

\newcommand{\Z}[1]{\ensuremath{\mathbbm{Z}_{#1}}}
\newcommand{\rep}[1]{\ensuremath{\mathbf{#1}}}

\newcommand{\U}[1]{\ensuremath{\mathrm{U}(#1)}}
\newcommand{\SO}[1]{\ensuremath{\mathrm{SO}(#1)}}

\newcommand{\E}[1]{\ensuremath{\mathrm{E}_{#1}}}

\newcommand{\D}{\mathrm{d}}
\newcommand{\mean}[1]{\ensuremath{\left\langle #1\right\rangle}}

\newcommand{\N}{\ensuremath{\mathcal{N}}}
\newcommand{\SGE}[2]{\ensuremath{\ensuremath{\left(#1,\,#2\right)}}}

\title{\boldmath String scale interacting dark matter from $\pi_1$}
\author{Andreas M\"utter}
\author{and Patrick K.S.~Vaudrevange}
\affiliation{Physics Department T75, Technical University of Munich\\
 James--Franck--Str.~1, 85748 Garching}

\emailAdd{andreas.muetter@tum.de}
\emailAdd{patrick.vaudrevange@tum.de}

\abstract{We show that in a wide class of string derived models of particle physics, heavy string 
modes with masses around the GUT scale can serve as a viable dark matter candidate. These heavy 
string modes wind around specific cycles in the extra-dimensional space, closely related to the 
fundamental group $\pi_1$. As a consequence of a non-trivial $\pi_1$, there is an exact discrete 
symmetry that stabilizes such winding strings. The dark matter candidate couples to the Standard 
Model via gravity and via the exchange of heavy string states. We find that, for reasonable values 
of the string coupling, our dark matter candidate can be produced in sizable amounts via freeze-in. 
Our scheme applies to many string constructions, including Calabi--Yau compactifications, and can 
be tested against constraints from the CMB.}

\arxivnumber{1912.09909}

\begin{document}
\maketitle

\section{Introduction}
Successful models for particle dark matter consist of two main ingredients: the dark matter 
candidate that interacts with the Standard Model (SM) only very weakly, and a mechanism that 
ensures its relative longevity. In most instances, a \Z2 symmetry is invoked to keep the dark 
matter particle from decaying. Moreover, if the model is supposed to explain the presently observed 
dark matter relic density, one needs to make sure that it is produced in sufficient quantities in 
the early universe. The prime example for a dark matter candidate is the weakly interacting massive 
particle (WIMP), which is in thermal equilibrium with the thermal plasma and ``freezes out'' after 
dropping out of equilibrium~\cite{Kolb:1990vq, Jungman:1995df}. However, it has been demonstrated 
that a dark matter species may also be produced thermally in sizable quantities even if it never 
attains thermal equilibrium (``freeze-in'') \cite{Hall:2009bx}. More specifically, it has become 
clear that freeze-in production can work with extremely heavy dark matter candidates (with masses 
above the GUT scale) and couplings that are suppressed by $1/m_\mathrm{Pl}^2$, a framework that is 
known as Planckian interacting dark matter (PIDM)~\cite{Garny:2015sjg,Garny:2017kha}. This 
observation is intriguing from the viewpoint of string model building, for a number of reasons. On 
the one hand, it is expected that the lightest massive string states have masses around the string 
scale, and that there are, apart from gravity, stringy interactions between these states and the 
massless states of the Standard Model that are suppressed by $1/m_\mathrm{s}^2$. Furthermore, there 
is a wide class of string models that have a stabilizing symmetry built in~\cite{Kogan:1986fx}, 
e.g.~an abelian \Z2 symmetry \cite{Ramos-Sanchez:2018edc}. The goal of this paper is to show that 
this class of string models can yield viable models of dark matter. As an explicit example, we 
consider the framework of heterotic string theory with six extra dimensions compactified on a 
special class of orbifolds but our scheme is valid more generally. In particular, our dark matter 
candidate is given by a closed string state that winds around a certain cycle in the 
extra-dimensional compact space. Then, its stability can be ensured via an exact discrete symmetry 
that originates from the topological property of certain compactification spaces to be non-simply 
connected. Consequently, the discrete symmetry is classically exact and can potentially be broken 
only non-perturbatively due to a discrete anomaly~\cite{Banks:1991xj,Lee:2011dya}. This stable 
string state is generically very heavy (with a mass at the compactification or GUT scale) and 
interacts with the massless states in the plasma not only via gravity, but also via the exchange of 
other heavy winding modes.

\begin{figure}[t!]
\centering
\begin{minipage}{0.49\textwidth}
\centering
\includegraphics[width=.8\textwidth]{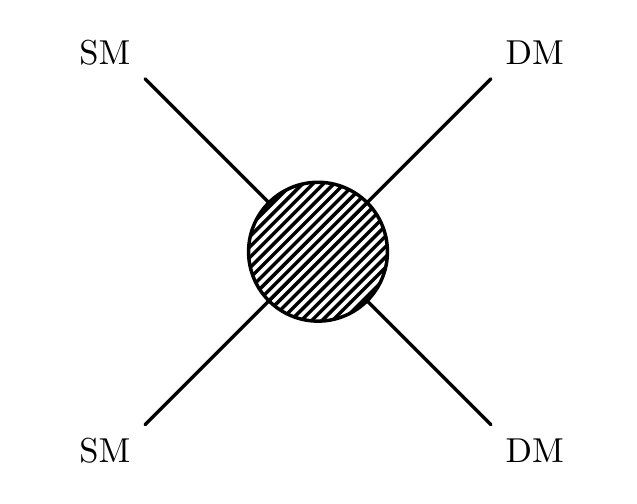}
\end{minipage}
\hfill
\begin{minipage}{0.49\textwidth}
\centering
\includegraphics[width=.8\textwidth]{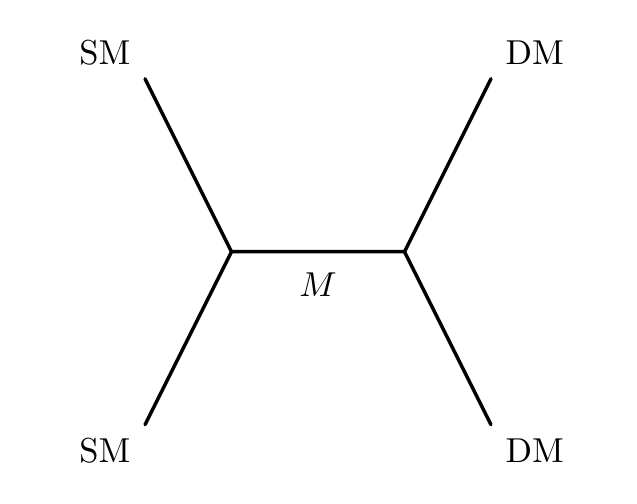}
\end{minipage}
\vspace{-0.4cm}
\caption{Dominant dark matter production by the exchange of a mediator field $M$. Note that the 
dark matter particle is stabilized by an odd $\Z{4}$ charge, while Standard Model fields carry even 
$\Z{4}$ charges and the mediator is uncharged. Importantly, the existence of such a stabilizing 
discrete symmetry is a direct consequence from the extra dimensions of string theory.}
\label{Fig:DMscatter}
\end{figure}

\section{Interactions between dark matter and the Standard Model}
In this section, we will examine which couplings between the dark matter candidate and the 
particles of the Standard Model can arise at the \emph{renormalizable} level. To be specific, 
we consider a string model compactified on the so-called $\Z2\times\Z2$-5-1 orbifold in the 
classification of ref.~\cite{Fischer:2012qj}. However, our findings easily carry over to 
other orbifold geometries and also to other string compactifications, cf.~appendix~\ref{App:StringyRealization} 
for further details. In this model, the dark matter candidate is stabilized by an exact \Z 4 symmetry 
that originates from string selection rules~\cite{Ramos-Sanchez:2018edc}, which in turn are 
related to topological properties of the compact orbifold space. It should be noted that all 
massless strings, especially those for the Standard Model particles, have even \Z 4 charges, while 
the DM particles carry odd charges. Hence, the \Z 4 has a \Z 2 subgroup with precisely the charge 
assignment needed for dark matter parity and the lightest string with odd \Z{2} charge is stable. Now, 
one has to examine the model further, in order to identify all allowed stringy couplings of the DM 
particle to the Standard Model. Since we are dealing with a supersymmetric setup, one needs to 
distinguish between couplings from the K\"ahler potential and those from the superpotential. 

In most instances, the coupling of dark matter to the SM is dominated by $2\rightarrow 2$ scattering. 
Hence, due to the constraining \Z 4 symmetry, the most general coupling looks like the left diagram 
in figure~\ref{Fig:DMscatter}. At tree level, one can boil that down to the exchange of mediator 
fields $M$, cf.~the right diagram in figure~\ref{Fig:DMscatter}. What are possible mediator fields? 
One example are gravitational interactions, where the exchanged particle is the graviton. This case 
has been studied extensively in the PIDM program~\cite{Garny:2015sjg, Garny:2017kha} and, as we 
will see, will give in our setup a contribution that is in general subdominant. However, it turns 
out that once interactions from string theory are considered, there are additional stringy 
mediators that can be exchanged and dominate the coupling between DM and the SM. These mediators 
are also winding strings but with vanishing \Z 4 charges. Hence, at generic points in string 
moduli space they are very massive. On the other hand, they have the generic feature to couple to 
the winding dark matter candidate and to Standard Model matter under the assumption that the SM 
matter is localized appropriately in the extra-dimensional orbifold space, cf.~appendix~\ref{App:StringyRealization}. 
The exact realizations of winding dark matter, winding mediators and localized SM matter fields 
depend on the specific string model. However, instead of discussing a full string realization, we 
consider a very generic string setting as presented in table~\ref{Tab:FieldsAndCharges}. 
Consequently, our findings will be rather generic for a wide class of string constructions.

\begin{table}[t!]
\centering
\begin{tabular}{c|lcc}
                                          & superfield                             & type of closed string & \Z 4 charge \\
\hline
\rule{0pt}{3ex}                  SM       & $\Phi_i=(f_i, \tilde{f}_i)$            & localized       & 0 or 2  \\
\hline
\rule{0pt}{3ex} \multirow{2}{*}{DM}       & $\Phi_\mathrm{DM}=(\chi, \varphi)$     & $\tau$-winding  & 1  \\
\rule{0pt}{3ex}                           & $\Phi_\mathrm{DM}'=(\chi', \varphi')$  & $-\tau$-winding & 3  \\
\hline
\rule{0pt}{3ex} \multirow{2}{*}{mediator} & $\Phi_\mathrm{M}=(\chi_\mathrm{M}, M)$ & winding         & 0  \\
\rule{0pt}{3ex}                           & $V^\mathrm{(M)}=(V_\mu, \lambda)$      & winding         & 0  \\
\end{tabular}
\caption{Summary of the relevant fields for SM, DM and mediator and their corresponding types of 
strings (i.e.\ localized strings, $\tau$-winding strings or general winding strings). 
$\Phi_\mathrm{DM}'$ denotes the mass partner of the dark matter multiplet $\Phi_\mathrm{DM}$.}
\label{Tab:FieldsAndCharges}
\end{table}

\subsection{K\"ahler potential terms}

If (at least part of) the massless Standard Model matter is localized in the extra-dimensional 
compact space, there is a set of mediators that can couple to both, the dark matter candidate and 
the Standard Model. In our concrete setting, there are three types of mediator strings, where each 
of them potentially couples to a different subset of SM matter. In what follows, we will make the 
simplifying assumption that all three have the same mass and couplings,\footnote{This assumption 
stems from the fact that the couplings of the winding mediator to localized strings is either of 
the same order (when both localized strings live at the same point in extra dimensions), or 
suppressed exponentially with their distance in extra dimensions.} such that one can effectively 
work with one mediator $V^\mathrm{(M)}$. As we show in appendix~\ref{sec:NarainDiscussion}, the 
mediator field originates from a massive string state that necessarily carries both Kaluza--Klein 
momentum and winding. Moreover, we shall see later on that variations of the coupling 
strengths do not have a large impact on our results. Let us now examine the coupling of dark 
matter to a localized Standard Model particle $\Phi_i$. Then, the most general K\"ahler potential 
consistent with gauge invariance and the stringy $\Z{4}$ symmetry for the coupling of the Standard 
Model to the mediator reads
\begin{align}
K_\mathrm{SM} ~\supset~\Phi_i^\dagger\left[\mathrm{e}^{2g_1V^\mathrm{(M)}}+\frac{g_1'}{\Lambda}\left(\Phi_\mathrm{M} + \Phi_\mathrm{M}^\dagger\right)+\frac{\I g_1''}{\Lambda}\left(\Phi_\mathrm{M} - \Phi_\mathrm{M}^\dagger\right)\right]\Phi_i\;.
\end{align}
Here, we observe that only the coupling to the vector field $V^\mathrm{(M)}=(V_\mu,\lambda)$ with 
coupling $g_1$ is renormalizable, hence all other terms will be dropped. Therefore, we consider 
only the coupling of the mediator vector superfield $V^{(\mathrm{M})}$ to the dark matter 
candidates $\Phi_\mathrm{DM}$ and $\Phi_\mathrm{DM}'$ in the dark matter K\"ahler potential
\begin{align}\label{eq:KahlerPot}
K_\mathrm{DM} ~\supset~ \Phi_\mathrm{DM}^\dagger\,\mathrm{e}^{2g_2V^\mathrm{(M)}}\Phi_\mathrm{DM} + \Phi_\mathrm{DM}'^\dagger\,\mathrm{e}^{-2g_2V^\mathrm{(M)}}\Phi_\mathrm{DM}'\;,
\end{align}
see appendix~\ref{sec:NarainDiscussion} for the stringy origin of these couplings. Let us 
parameterize the SM chiral multiplets as $\Phi_i=(f_i,\tilde{f}_i)$ and the dark matter multiplets 
as $\Phi_\mathrm{DM}=(\chi,\varphi)$ and $\Phi_\mathrm{DM}'=(\chi',\varphi')$. The relevant 
Lagrangian for the $2\rightarrow 2$ production of dark matter from the $D$-terms of the K\"ahler 
potentials reads
\begin{align}
\mathcal{L}~\supset&~K_\mathrm{SM}\left.\vphantom{\int}\!\right\vert_D + K_\mathrm{DM}\left.\vphantom{\int}\!\right\vert_D\\[1ex]
~\supset&~ g_1 V^{(\mathrm{M})}_\mu\left[\left(\overline{f}_i\overline{\sigma}^\mu f_i+2\I\tilde{f}_i^\dagger\partial_\mu\tilde{f}_i\right)+ g_2\left(\overline{\chi}\,\overline{\sigma}^\mu\chi+2\I\varphi^\dagger\partial_\mu\varphi\right)\right]\nonumber\\[1ex]
&+\sqrt{2}g_1\left(\tilde{f}_i\overline{\lambda}\,\overline{f}_i + \tilde{f}_i^\dagger\lambda f_i \right)+\sqrt{2}g_2\left(\varphi\overline{\lambda}\overline{\chi} + \varphi^\dagger\lambda\chi \right) + \begin{pmatrix}\chi\leftrightarrow\chi'\\\varphi\leftrightarrow\varphi'\\g_2\leftrightarrow -g_2\end{pmatrix}\;.
\end{align}

\begin{figure}[t!]
\centering
\begin{minipage}{0.3\textwidth}
\includegraphics[width=\textwidth]{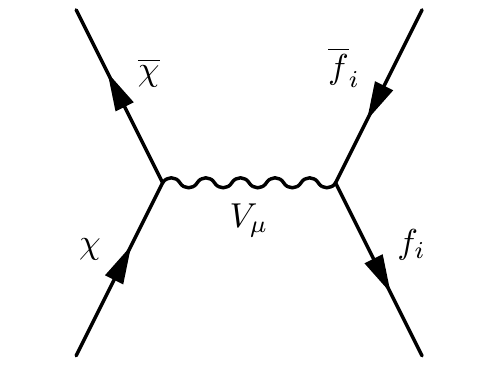}
\captionof{figure}{$\chi\overline{\chi}\leftrightarrow f_i\overline{f}_i$\label{Fig:Scatter1}}
\end{minipage}\hfill
\begin{minipage}{0.3\textwidth}
\includegraphics[width=\textwidth]{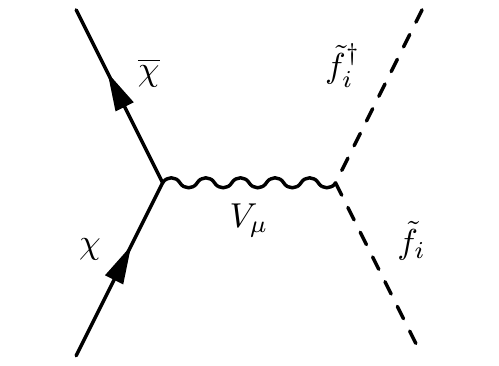}
\captionof{figure}{$\chi\overline{\chi}\leftrightarrow \tilde{f}_i^\dagger\tilde{f}_i$}
\end{minipage}\hfill
\begin{minipage}{0.3\textwidth}
\includegraphics[width=\textwidth]{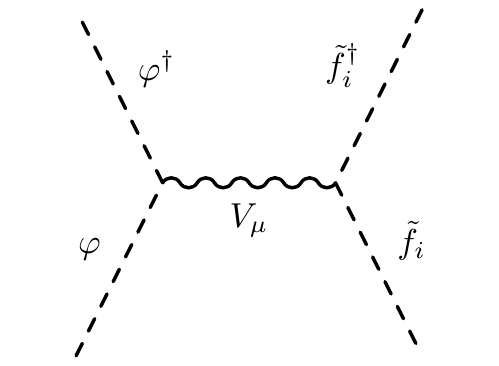}
\captionof{figure}{$\varphi^\dagger\varphi\leftrightarrow\tilde{f}_i^\dagger\tilde{f}_i$}
\end{minipage}
\end{figure}

\begin{figure}[t!]
\centering
\begin{minipage}{0.3\textwidth}
\includegraphics[width=\textwidth]{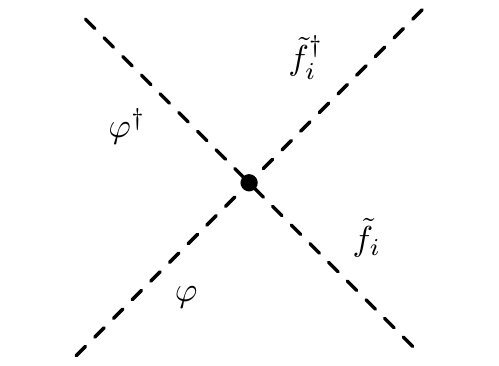}
\captionof{figure}{$\varphi^\dagger\varphi\leftrightarrow\tilde{f}_i^\dagger\tilde{f}_i$}
\end{minipage}\hfill
\begin{minipage}{0.3\textwidth}
\includegraphics[width=\textwidth]{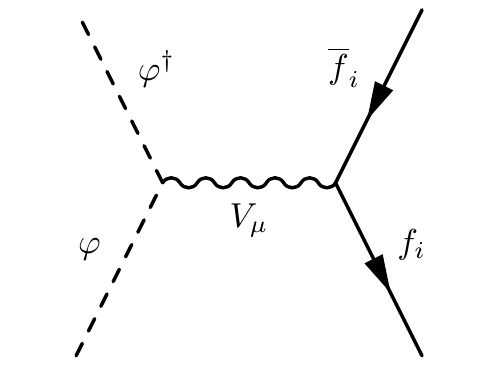}
\captionof{figure}{$\varphi^\dagger\varphi\leftrightarrow f_i\overline{f}_i$}
\end{minipage}
\hfill
\begin{minipage}{0.3\textwidth}
\includegraphics[width=\textwidth]{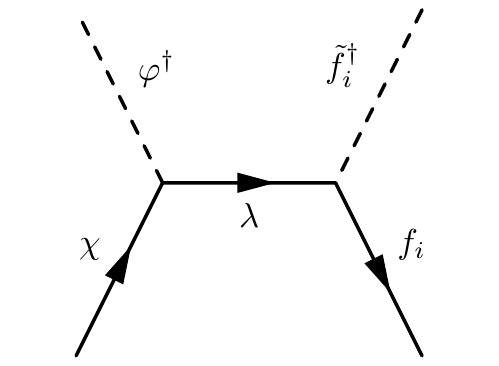}
\captionof{figure}{$\varphi^\dagger\chi\leftrightarrow\tilde{f}_i^\dagger f_i$\label{Fig:Scatter5}}
\end{minipage}
\end{figure}

Additionally, there is a four-scalar vertex coming from the kinetic term of the mediator multiplet. 
The Lagrangian coming from the auxiliary field in $V^{(\mathrm M)}$ reads
\begin{align}
\mathcal{L}_{(D_{\mathrm M})}~=~\frac{1}{2}D_{\mathrm M}^2 + g_1 D_{\text M}|\tilde{f}_i|^2 + g_2 D_{\mathrm M}|\varphi|^2 - g_2 D_{\mathrm M}|\varphi'|^2 + \dots\;,
\end{align}
which, upon setting the auxiliary field $D_\mathrm{M}$ on-shell, yields
\begin{align}
\mathcal{L}_{(D_{\mathrm M})}~=~-\frac{1}{2}\left(g_1|\tilde{f}_i|^2+g_2|\varphi|^2-g_2|\varphi'|^2\right)^2+\dots\;,
\end{align}
and, hence, we obtain a four-scalar vertex with a coupling $g_1g_2$. Then, we find that at tree level the 
relevant channels for the non-gravitational interactions of dark matter with the Standard 
Model are given by the processes shown in figures~\ref{Fig:Scatter1}--\ref{Fig:Scatter5}. There, we 
present the production channels for the dark matter multiplet $\Phi_\mathrm{DM}$, analogous diagrams 
exist also for its partner multiplet $\Phi_\mathrm{DM}'$. 

Let us briefly discuss the conceivable range of values for the couplings $g_1$ and $g_2$. In 
supersymmetric gauge theories, each gauge coupling is given by a gauge kinetic function $f$. For 
example, in the case of the $\U{1}$ associated with the mediator field $V_\mu^{(\text{M})}$ we have
\begin{align}
f_{\U{1}} ~=~ S + \Delta_{\U{1}}( T_i, U_i)\;,
\end{align}
where $S$ is the heterotic axio-dilaton and the threshold correction $\Delta_{\U{1}}$ is a stringy 
one-loop contribution that is in general a complicated function of the geometric moduli $T_i$ and 
$U_i$, see ref.~\cite{Dixon:1990pc}. However, for the non-factorizable orbifold we are considering, 
the precise form of $\Delta_{\U{1}}$ is unknown. Still, we expect that 
by varying the geometric moduli, one can generate wide ranges of effective couplings for the 
mediator \U1. The couplings of the Standard Model gauge group follow a similar pattern. However, 
the threshold corrections $\Delta_{\mathrm{SM}}$ for the Standard Model have in general a different 
dependence on the geometric moduli than $\Delta_{\U{1}}$. Hence, it is conceivable that the 
mediator couplings can be varied without spoiling the unification of the Standard Model gauge 
couplings.

\subsection{Superpotential terms} In addition to the K\"ahler terms, there can also arise couplings 
from the superpotential. In particular, it is possible to couple a mediator, residing in a chiral 
multiplet, to the Standard Model via Higgs portal and neutrino-portal-like terms. If we make the 
assumption that the mediator $\Phi_\mathrm{M}$ couples to all three generations of lepton doublets 
$L$ with the same coupling constant $\lambda_N$, the terms containing the mediator in the 
corresponding superpotential read
\begin{align}
\mathcal{W}~=&~\mathcal{W}_\mathrm{M} + \mathcal{W}_\mathrm{DM} + \mathcal{W}_\mathrm{Higgs-portal}+ \mathcal{W}_\mathrm{neutrino-portal}
\\[1ex]
~=&~\frac{m_\mathrm{M}}{2}\,\Phi_\mathrm{M}^2+\frac{\lambda_\mathrm{M}}{3}\,\Phi_\mathrm{M}^3
+
\frac{m_\mathrm{DM}}{2}\,\Phi_\mathrm{DM}\,\Phi_\mathrm{DM}'
+
\lambda_\mathrm{DM}\,\Phi_\mathrm{M}\,\Phi_\mathrm{DM}\,\Phi_\mathrm{DM}'
\notag\\[1ex]
&~+
\lambda_\mathrm{H}\,\Phi_\mathrm{M}\,\widehat H_u\,\widehat H_d
+
\lambda_\mathrm{N}\,\Phi_\mathrm{M}\,\widehat H_u\,\widehat L\;,
\end{align}
where we used the SM superfields $\widehat H_{u,d}=(\widetilde H_{u,d},  H_{u,d})$ for the Higgses 
and $\widehat L=(\ell, \tilde{\ell})$ for the lepton doublet(s). For processes involving the 
exchange of a mediator, we are interested in the 3-point interactions that arise from this choice 
for $\mathcal{W}$. Additionally, there is also a four-scalar interaction of two dark matter scalars 
and two SM scalars. However, in order for the stringy couplings to exist, the Higgs field must be 
localized.  For the neutrino portal, the Higgs field has to live at the same fixed point as the 
lepton doublet it is supposed to couple to; if these fields live at different fixed points in the 
same sector the couplings are suppressed, and if they live in another twisted sector the couplings 
are completely absent. Similar terms then exist for the coupling of the mediator to the dark matter 
candidate. It turns out that if the couplings $\lambda_H$ and $\lambda_N$ are chosen to be of the same order 
as the K\"ahler couplings $g_{1,2}$, the contribution of the superpotential couplings to the dark 
matter production rate is not qualitatively different from the contribution of the K\"ahler 
potential, but numerically a little bit lower. Moreover, as discussed above, the existence of 
the neutrino portal couplings requires an specific localization of the lepton and Higgs fields, 
which is model dependent. For these reasons, and because models with localized Higgs pairs have a 
less appealing phenomenology, we will not push any further in this direction and consider K\"ahler 
terms only by assuming the Higgs field to originate from the bulk.

\section{Dark matter production}
\begin{figure}[t!]
\centering
\includegraphics[width=0.65\textwidth]{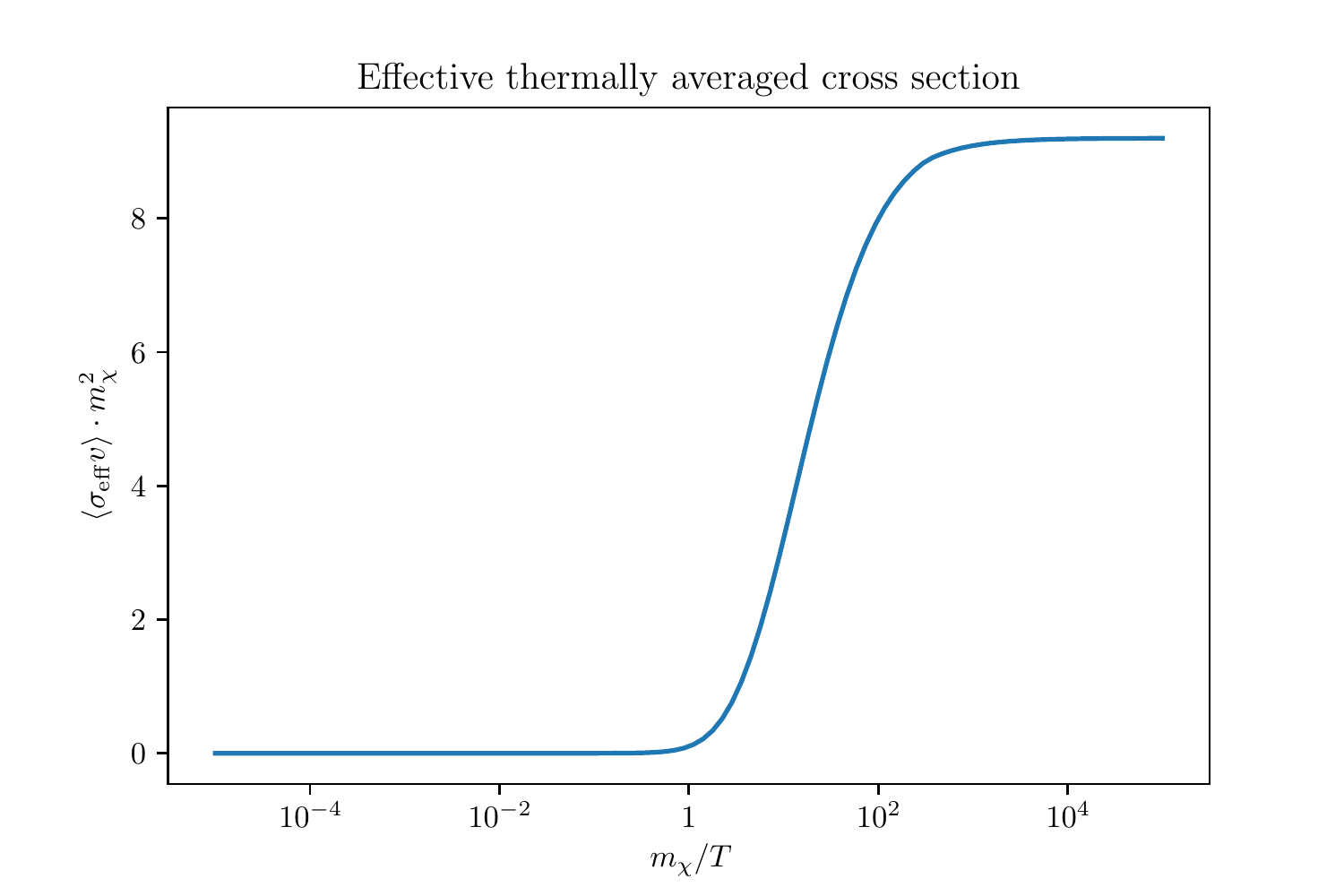}
\vspace{-0.4cm}
\caption{The effective thermally averaged cross section for a mediator mass $m_\mathrm{M}=1.8\,m_\chi$ and all couplings set to unity. Observe how the cross section approaches a constant value for $T \ll m_\chi$.}
\label{Fig:ThermAvgKahler}
\end{figure}

Although the dark matter candidate is too heavy to be in thermal equilibrium, it can still be 
produced thermally via freeze-in. The production of dark matter is then governed by the Boltzmann equation
\begin{align}\label{Eq:Boltzmann}
\dot{n}+3H(t)\,n~=~-\mean{\sigma_\mathrm{eff}v}\left(n^2-n_\mathrm{eq}^2\right)\;.
\end{align}
Here, $n$ is the number density of all states in the dark matter sector, in other words 
$n=n_\chi+n_\varphi+n_{\overline{\chi}}+n_{\varphi^\dagger}+n_{\chi'}+n_{\varphi'}+n_{\overline{\chi}'}+n_{\varphi^{'\dagger}}$. 
On the right hand side of equation~\eqref{Eq:Boltzmann}, $\mean{\sigma_\mathrm{eff}v}$ is the 
effective thermally averaged cross section for the various $2\rightarrow 2$ dark matter production 
channels, taking also coannihilations into account~\cite{Gondolo:1990dk, Edsjo:1997bg}. Using 
$m_i = m_\chi$ it reads
\begin{align}\label{Eq:ThermalAv}
\mean{\sigma_\mathrm{eff}v}~=~\frac{T}{n_\mathrm{eq}^2}\frac{1}{8\pi^4}
\int_{4m_\chi^2}^\infty \D s \sqrt{s}\,p^2\,\left(\sum_{i,j}g_ig_j\sigma_{ij}(s)\right)\, K_1\left(\frac{\sqrt{s}}{T}\right)\;.
\end{align}
Here, $g_i$ counts the internal degrees of freedom of each species $i$ (where $g_i=2$ for a Weyl 
fermion and $g_i=1$ for a real scalar), and the summation indices $i$ and $j$ run over all 
fields in the dark sector. Furthermore, $p=\sqrt{s/4-m_\chi^2}$, and $K_1$ is the modified Bessel 
function of the second kind of order 1. The equilibrium density $n_\mathrm{eq}$ is given by
\begin{align}
n_\mathrm{eq}~=~\sum_i\frac{T}{2\pi^2}g_i\,m_i^2K_2\left(\frac{m_i}{T}\right)~=~\frac{4T}{\pi^2}\,m_\chi^2K_2\left(\frac{m_\chi}{T}\right)\;,
\end{align}
where $K_2$ is the modified Bessel function of the second kind of order 2. The cross sections 
$\sigma_{ij}$ correspond to the various possible scattering processes shown in 
figures~\ref{Fig:Scatter1}--\ref{Fig:Scatter5}, and are given by
\begin{align}
\sigma_{ij}~=~\frac{1}{16\pi s\left(s-4m_\chi^2\right)}\int_{t_-}^{t_+}\D t\left|\mathcal{M}_{ij}(t)\right|^2\;.
\end{align}
Here, $t_\pm=-\left(\sqrt{s/4}\mp\sqrt{s/4-m_\chi^2}\right)^2$ and $\mathcal{M}_{ij}(t)$ denote the 
matrix elements for the respective process. In what follows, we focus on the case with 
bulk Higgs fields and, hence, there are no contributions from the superpotential. Then, the 
non-vanishing cross sections $\sigma_{ij}$ are given by
\begin{align}
\sigma_{\chi\overline{\chi}}~=&~\sigma_{\chi\overline{\chi}\rightarrow f_i\overline{f}_i}+\sigma_{\chi\overline{\chi}\rightarrow \tilde{f}_i\tilde{f}_i^\dagger}\\
\sigma_{\varphi\varphi^\dagger}~=&~\sigma_{\varphi\varphi^\dagger\rightarrow f_i\overline{f}_i}+\sigma_{\varphi\varphi^\dagger\rightarrow\tilde{f}_i\tilde{f}_i^\dagger}\\
\sigma_{\chi\varphi^\dagger}~=&~\sigma_{\chi\varphi^\dagger\rightarrow f_i \tilde{f}_i^\dagger}\\
\sigma_{\overline{\chi}\varphi}~=&~\sigma_{\chi\varphi^\dagger}\;,
\end{align}
plus the corresponding terms for $\chi',\varphi'$. Moreover, it holds that $\sigma_{ij}=\sigma_{ji}$. 
With these preparations in place, one can perform the thermal averaging eq.~\eqref{Eq:ThermalAv} numerically 
(cf.~figure~\ref{Fig:ThermAvgKahler}) and turn one's attention to the Boltzmann equation.
As the actual density in the freeze-in case is always much smaller than the equilibrium one, the 
full Boltzmann equation~\eqref{Eq:Boltzmann} can be approximated to sufficient accuracy by 
neglecting $n^2$ compared to $n_\mathrm{eq}^2$ on the right hand side of equation~\eqref{Eq:Boltzmann} 
and hence using
\begin{align}\label{Eq:BoltzmannFreezeIn}
\dot{n}+3H(t)\,n~=~\mean{\sigma_\mathrm{eff}v}\,n_\mathrm{eq}^2\;.
\end{align}
Proceeding like in ref.~\cite{Garny:2015sjg}, one can now simplify the discussion by 
introducing the dimensionless abundance $X=na^3/T_\mathrm{rh}^3$ in terms of the scale factor $a$ and 
the reheating temperature $T_\mathrm{rh}$, such that equation~\eqref{Eq:BoltzmannFreezeIn} can be 
integrated to yield
\begin{align}\label{Eq:FinalAbundance}
X_\mathrm{f}~=~\frac{1}{T_\mathrm{rh}^3}\int_1^\infty\D a\frac{a^2}{H(a)}\mean{\sigma_\mathrm{eff}v}\,n_\mathrm{eq}^2\;.
\end{align}
Here, we used the fact that the scale factor at the end of inflation can be chosen to be 1, and 
that the abundance of dark matter immediately after inflation vanishes. The maximal possible relic 
abundance is obtained if the reheating phase after inflation is as short as possible, leading to 
the highest maximal temperature that is reached during reheating. Scenarios with this instantaneous 
reheating require
\begin{align}
\frac{H_i}{\Gamma}~\sim~1\;,
\end{align}
where $H_i$ is the Hubble rate at the end of inflation and $\Gamma$ the inflaton decay rate. Then, 
the reheating temperature coincides with the highest temperature reached during reheating and is 
given by
\begin{align}
T_\mathrm{rh}~\approx~ 0.25\sqrt{m_\mathrm{Pl}\,H_i}\;.
\end{align}
While non-perturbative reheating scenarios \cite{Felder:1998vq} provide a straightforward way to 
achieve this, they also imply the non-thermal production of (heavy) particles, as opposed to 
perturbative scenarios of reheating. However, it has been shown that one can realize a 
near-instantaneous reheating scenario also within the context of perturbative reheating \cite{Garny:2017kha}, 
which we will also assume throughout this work. By doing so, we obtain an upper limit on the 
amount of thermally produced dark matter for a given Hubble rate $H_i$. Equivalently, this can be 
seen as a lower bound on the Hubble rate $H_i$ needed in order to explain the observed relic 
density $\Omega_Xh^2$ by our dark matter candidate only. On the other hand, the non-observation of 
tensor modes in the cosmic microwave background (CMB) by the Planck satellite combined with constraints 
from \textsc{Bicep}2 and Keck requires a tensor-to-scalar ratio $r < 0.056$~\cite{Akrami:2018odb}. 
This gives an upper limit on $H_i$ and therefore on the reheating temperature
\begin{align}
T_\mathrm{rh}~<~5.8\cdot 10^{-4} m_\mathrm{Pl}~\approx~7\cdot 10^{15}\,\mathrm{GeV}\;.
\end{align}
Note that this bound is believed to become more stringent in the near future \cite{Errard:2015cxa}. 
Upon adopting the convention that the scale factor after inflation $a_i=1$, the dependence of the 
temperature and the Hubble rate on the scale factor for the radiation dominated phase after reheating is
\begin{align}
T(a)~=~\frac{T_\mathrm{rh}}{a}\;,\quad H(a)~=~\frac{H_i}{a^2}\;.
\end{align}
Thus, the abundance eq.~\eqref{Eq:FinalAbundance} can be seen as a function 
$X_\mathrm{f}(H_i, g_1g_2, m_\chi, m_\mathrm{M})$ of the Hubble rate at the end of inflation $H_i$, 
the couplings $g_1g_2$, the dark matter mass $m_\chi$ and the mediator mass $m_\mathrm{M}$.
In order to compare to the observed dark matter relic density $\Omega_X h^2 = 0.12$~\cite{Aghanim:2018eyx}, 
one can use (cf.~\cite{Garny:2017kha})
\begin{align}\label{Eq:Xcrit}
X_\mathrm{f}^\mathrm{crit.}~=~ 0.29\cdot 10^{-5}\cdot \frac{\mathrm{GeV}}{m_\chi}\cdot \Omega_X h^2 \;.
\end{align}
Hence, for a GUT scale dark matter particle ($m_\chi\sim 10^{16}\,\mathrm{GeV}$), the critical 
abundance is of order $10^{-23}$. It is interesting to notice that the Hubble rate $H_i$ required 
to obtain this abundance remains relatively stable even if vectorlike SM exotics are added, owing 
to the nature of freeze-in production. To see this, note that if the couplings of all contributing 
chiral multiplets are roughly equal, the Hubble rate needed to match the correct final abundance is 
determined by the contribution $x_\mathrm{f} = X_\mathrm{f}/N_\Phi$ of a single multiplet to the final 
abundance. Then, the critical contribution per chiral multiplet be written as
\begin{align}
x_\mathrm{f}^\mathrm{crit.}~=~\frac{g_\ast}{N_\Phi}\,R\;,
\end{align}
where $g_\ast$ counts the number of degrees of freedom in the thermal bath at $T_\mathrm{rh}$ and 
$N_\Phi$ is the number of contributing chiral multiplets (for the case of the MSSM with three right 
handed neutrinos, $g_\ast=240$ and $N_\Phi=48$). Adding $n_V$ vectorlike pairs of exotics now 
changes these figures according to
\begin{equation}
g_\ast ~\mapsto~ g_\ast+7.5\,n_V \quad\mathrm{and}\quad N_\Phi~\mapsto~ N_\Phi+2\,n_V\;.
\end{equation}
Hence, adding an arbitrary number of vectorlike exotics lowers $x_\mathrm{f}^\mathrm{crit.}$ by at 
most 25\%. This change requires an even smaller adjustment in the Hubble rate $H_i$, and therefore 
our results are largely insensitive to the full particle content of a given model.

\begin{figure}[t!]
\begin{minipage}{0.49\textwidth}
\centering
\includegraphics[width=1.0\textwidth]{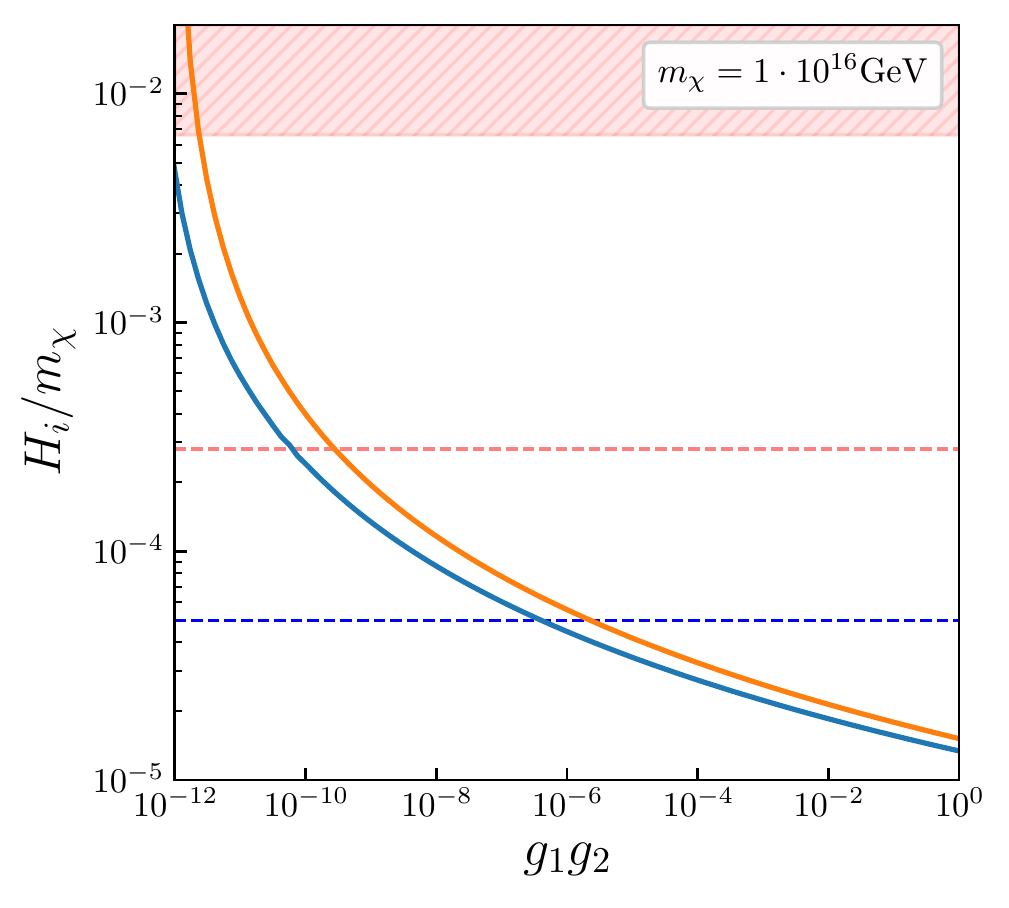}
\end{minipage}
\hfill
\begin{minipage}{0.49\textwidth}
\centering
\includegraphics[width=1.0\textwidth]{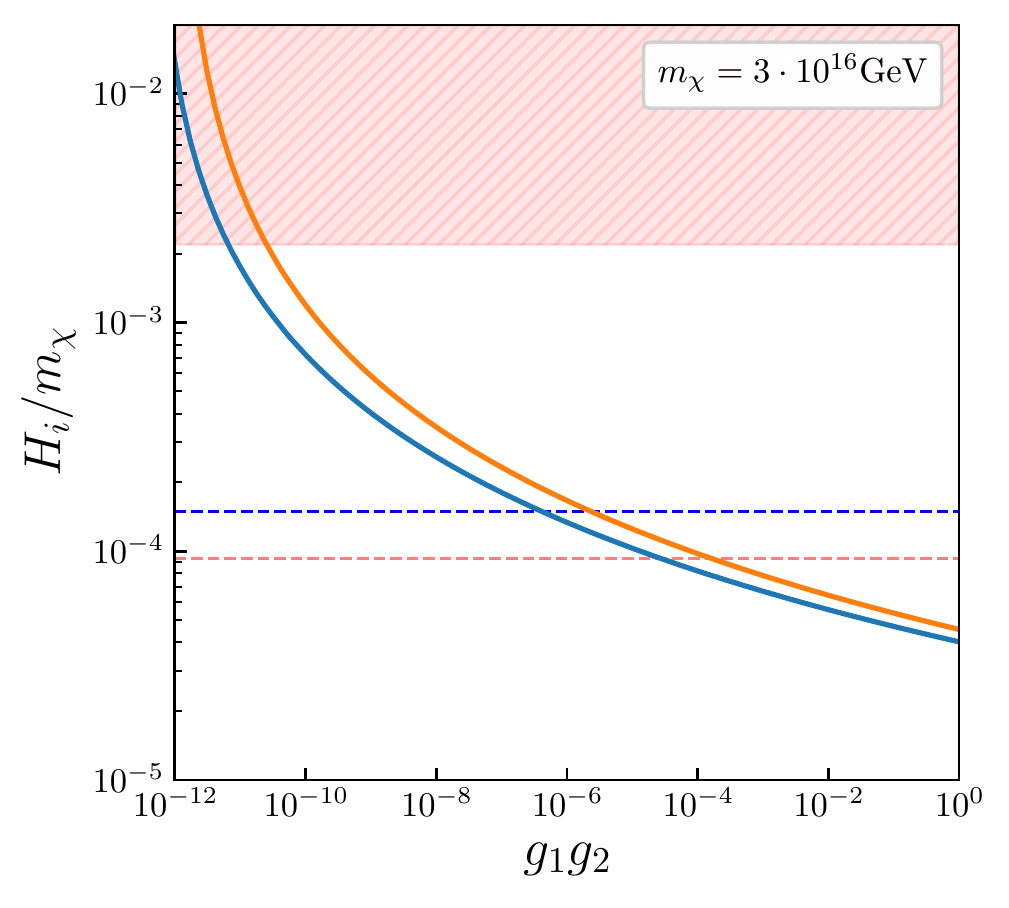}
\end{minipage}
\vspace{-0.4cm}
\caption{Plot of the critical Hubble rate at the end of inflation as a function of the product of the couplings, for a dark matter mass of $10^{16}\mathrm{ GeV}$ (left) and $3\cdot 10^{16}\mathrm{ GeV}$ (right). The blue (orange) curve indicates a mediator mass of $1.9\,m_\chi$ $(1.0\,m_\chi)$. The red area at the top is excluded by the currently observed bound for the tensor-to-scalar ratio in the CMB. Additionally, the projected sensitivity of CMB experiments is shown as the dashed red line.  Furthermore, the critical Hubble rate of \cite{Garny:2015sjg} is indicated by the blue dashed line.}
\label{Fig:ScanCouplings}
\end{figure} 

\section{Results}
We have solved the integral \eqref{Eq:FinalAbundance} numerically. If we use the simplified 
reheating scenario and a fixed value of the dark matter mass $m_\chi$, the resulting abundance 
$X_\mathrm{f}$ depends only on the Hubble rate $H_i$ after inflation (which sets the reheating 
temperature), and the product of the two involved couplings. In principle, there is also a light 
dependence on the mediator mass $m_\mathrm{M}$, however as one observes, varying the mediator mass 
shows only little effect on the final abundance, especially for larger values of the couplings. 
Our results are displayed in figure~\ref{Fig:ScanCouplings}. There, we varied the couplings 
$g_1g_2$ over a broad range, and determined the value of $H_i$ needed to produce the critical 
dark matter relic abundance $X_\mathrm{f}^\mathrm{crit.}$, eq.~\eqref{Eq:Xcrit}.

One observes that for very small couplings, values for the Hubble rate that exceed the current 
CMB bounds \cite{Akrami:2018odb} are needed in order to produce the right amount of dark matter. 
The bounds are more strict for lower mediator masses. Likewise, the critical Hubble rate changes 
for less than an order of magnitude for a wide range of coupling strengths, roughly from 
$g_1g_2=10^{-8}$ to $g_1g_2=1$.

Moreover, one observes that for values of $g_1g_2$ greater than $10^{-6}$, the production of dark 
matter via the stringy operators largely dominates the graviton exchange presented in 
ref.~\cite{Garny:2015sjg, Garny:2017kha} and therefore neglecting this gravitational channel is a 
good assumption. For lighter dark matter masses it is impossible to get near the CMB bound, even 
the projected ones, without encountering overproduction by graviton exchange first. However, 
starting from $m_\chi\sim 3\cdot 10^{16}\mathrm{GeV}$, our approach becomes sensitive to at least 
the projected CMB bound in ranges for the stringy couplings for which graviton exchange can still 
be safely neglected. We also observe that for any value of the couplings, our Hubble rate lies in 
ranges where -- given the DM mass -- gravitational production \cite{Chung:2001cb} can be neglected.

\section{Conclusions}
We have shown that generic string constructions can accommodate a candidate for dark matter. 
Opposed to other studies of dark matter in string theory (cf.~\cite{Chowdhury:2018tzw, Bhattacharyya:2018evo, Bernal:2018qlk}), 
we focused on the dark matter candidate, as e.g.~in \cite{Halverson:2016nfq,Halverson:2018olu}. 
Specifically, the dark matter candidate is a heavy string state with no charge under the Standard 
Model gauge group and a mass at or above the GUT scale. It is stabilized against decay by stringy 
selection rules, closely related to the topological property of a non-trivial fundamental group 
$\pi_1$ of the compactification space. Because of its high mass, the dark matter particle never 
attains thermal equilibrium, and therefore it must be produced by freeze-in rather than freeze-out. 
We find that generically, the dark matter candidate interacts with the thermal bath only via 
gravity, and by the exchange of heavy mediators arising in the massive string spectrum. For not 
too small string couplings the latter ones dominate over graviton exchange.

For definiteness, we considered an explicit model in heterotic orbifolds, but we believe that our 
results carry over very well to many other string constructions, for example in Calabi--Yau 
constructions with freely-acting Wilson lines~\cite{Braun:2005nv,Anderson:2011ns,Anderson:2014hia}. 
In our setup, we chose a small-radius limit, where winding strings are the lightest extra states, 
in order to identify the stringy selection rules. There exists a $T$-dual large-radius picture 
where the winding states are exchanged by Kaluza--Klein excitations. In our picture, the dark matter 
candidate is a string with winding around a particular non-contractible cycle on the orbifold, 
thereby ensuring its stability. An analysis on the level of the orbifold space group reveals that 
this winding state can couple to the Standard Model -- apart from gravity -- via the exchange of 
heavy winding strings that are SM singlets. Going to an $\N=1$ supersymmetric field theory, we 
identified the relevant terms for $2\rightarrow 2$ production of the heavy dark matter candidate 
from the K\"ahler potential and the superpotential. We find that in order to obtain the correct 
dark matter relic density, one needs values for the Hubble rate at the end of inflation that range 
up to $10^{12}$ GeV, if the string coupling is perturbative. This way, one is able to constrain the 
allowed parameter space of the model by bounds on the tensor-to-scalar ratio in the CMB.

We observe that our results generalize very well to generic string models: The most prominent 
influence comes only from the mass of the dark matter candidate itself, which is constrained to lie 
around the GUT scale. Other model-dependent parameters, such as the mass of the mediator contribute 
only at subleading order. Furthermore, the required Hubble rate after inflation remains within the 
same order of magnitude for the entire sensible range of string couplings. Finally, we observe that 
adding vectorlike matter that is charged under the SM and contributes to the production of dark 
matter, changes the critical final abundance only by a small amount, and leaves the required 
Hubble rate invariant up to the percent level.

\paragraph{Acknowledgements.} This work was supported by the Deutsche Forschungsgemeinschaft (SFB1258). 
We would like to thank Mathias Garny and Kai Urban for useful discussions.

\appendix
\section{An explicit realization in string theory}\label{App:StringyRealization}
\label{sec:StringRealization}

We consider the $\E{8}\times\E{8}$ heterotic string theory compactified on the $\Z2\times\Z2$-5-1 
orbifold geometry~\cite{Dixon:1985jw,Dixon:1986jc,Ibanez:1986tp}, see also 
refs.~\cite{Forste:2006wq,Donagi:2008xy,Fischer:2012qj} and refs.~\cite{Blaszczyk:2009in,Olguin-Trejo:2018wpw} 
for MSSM-like string models based on this orbifold geometry. This orbifold geometry can be 
constructed in three steps. First, one defines a factorized six-torus 
$\mathbb{T}^6=\mathbb{T}^2\times\mathbb{T}^2\times\mathbb{T}^2$ via a six-dimensional lattice that 
is spanned by six basis vectors $e_i$, $i=1,\ldots,6$. Then, this six-torus is orbifolded by 
$\Z2\times\Z2$ rotations $\theta \mathrel{\widehat{=}} (0,\nicefrac{1}{2},-\nicefrac{1}{2})$ and 
$\omega \mathrel{\widehat{=}} (\nicefrac{1}{2},0,-\nicefrac{1}{2})$, indicating the rotation angles 
in units of $2\pi$ in the three complex coordinates corresponding to the three two-tori 
$\mathbb{T}^2$. By doing so, one obtains the $\Z2\times\Z2$-1-1 orbifold geometry. Finally, one 
defines the shift
\begin{equation}\label{eq:tau}
\tau ~=~ \frac{1}{2}\left(e_2 + e_4 + e_6\right)\;.
\end{equation}
The resulting six-torus $\mathbb{T}^6$ spanned by $e_1,\ldots,e_5$ and $\tau$ is non-factorizable. 
It turns out that $\tau$ acts freely on the $\Z2\times\Z2$-1-1 orbifold, i.e.\ there is no point on 
the $\Z2\times\Z2$-1-1 orbifold that is invariant under a shift by $\tau$. Hence, $\tau$ is called 
freely-acting. By modding out the $\Z2\times\Z2$-1-1 orbifold by $\tau$, one obtains the 
$\Z2\times\Z2$-5-1 orbifold geometry.

\subsection{Strings on orbifolds}

Closed strings on orbifolds are characterized by their boundary conditions that specify which 
transformation is needed such that the string is closed. In more detail, for a string (i.e.\ a 
worldsheet boson) $X(\sigma_0,\sigma_1)$ as a function of worldsheet time and space coordinates 
$\sigma_0$ and $\sigma_1\in[0,1]$ the boundary condition reads
\begin{equation}\label{eq:bc}
X(\sigma_0, \sigma_1+1) ~=~ \theta^k\,\omega^\ell\,X(\sigma_0,\sigma_1) + n_i\,e_i + n_\tau\, \tau \quad\Leftrightarrow\quad g~=~(\theta^k\,\omega^\ell, n_i\,e_i + n_\tau\, \tau) ~\in~ S\;,
\end{equation}
where $k, \ell \in \{0,1\}$, $n_i\in\Z{}$, $n_\tau\in\{0,1\}$ and summation over $i=1,\ldots,6$ is 
implied. Strings with $k \neq 0$ or $\ell \neq 0$ are called twisted strings, in contrast to the 
case $k=\ell=0$ which gives rise to so-called untwisted strings. One can encode the boundary 
condition~\eqref{eq:bc} into group elements $g\in S$ of the so-called space group $S$. Then, $g$ is 
called the constructing element of the string~\eqref{eq:bc}. In more detail, since $X$ and $h\,X$ 
are identified on the orbifold for all $h \in S$, a string is actually characterized not only by 
the single constructing element $g\in S$ but by the conjugacy class $[g]=\{h\,g\,h^{-1}~|~h\in S\}$. 
If $g\in [g] \subset S$ has a fixed point, i.e.\ if there is a point $x_g$ such that 
$\theta^k\omega^\ell x_g + n_i\,e_i + n_\tau\, \tau = x_g$, the string with constructing element 
$g$ is localized in the extra dimensions at $x_g$. It is important to remark that the freely-acting 
nature of $\tau$ becomes evident by noticing that constructing elements with fixed points 
necessarily have $n_\tau=0$. Furthermore, strings with boundary conditions 
$(\Id, n_i\,e_i + n_\tau\, \tau)\in S$ live in the orbifold bulk. They are winding strings if 
$n_i \neq 0$ or $n_\tau\neq 0$, where the mass of a winding string is proportional to the radius 
and the winding number of its winding direction, as we will discuss later in 
appendix~\ref{sec:NarainDiscussion}. Hence, in general only bulk strings with constructing element 
$(\Id,0)$ are massless. 

The $\Z2\times\Z2$-5-1 orbifold geometry has the important property of having a cycle that generates 
a non-trivial fundamental group $\pi_1$~\cite{Dixon:1986jc} and, hence, renders the orbifold geometry 
non-simply connected. In fact, this cycle is generated by the freely-acting shift $\tau$. The 
existence of the freely-acting shift has two important consequences for our discussion:
\begin{enumerate}
\item There are heavy string modes with constructing elements $(\Id, n_\tau\tau)\in S$ that wind 
around the freely-acting $\tau$-direction and
\item There is an exact \Z{4} symmetry~\cite{Ramos-Sanchez:2018edc}, where a string with general 
constructing element eq.~\eqref{eq:bc} carries a discrete charge
\begin{align}\label{eq:Z4Charge}
Q ~=~ n_\tau+2(n_2+n_4+n_6) \;\;\mathrm{mod}\;\; 4 \quad\mathrm{such\ that}\quad Q ~\in~ [0,1,2,3]\;,
\end{align}
where $n_\tau\in\{0,1\}$ and $n_2, n_4, n_6 \in \Z{}$ are the integer winding numbers. It turns out 
that all massless strings (those from the bulk and those that are localized at orbifold fixed 
points) have $n_\tau=0$ and, therefore, carry even \Z{4} charges, while there exist massive strings 
with odd \Z{4} charges.
\end{enumerate}
Consequently, there exists a lightest winding string from the bulk with winding numbers $n_\tau=1$ 
and $n_i=0$, i.e.\ with constructing element $\SGE{\Id}{\tau}$, that has odd $\Z{4}$ charge. 
Hence, it is stable and we can identify it as our dark matter candidate $\Phi_\mathrm{DM}$. Its mass 
partner $\Phi_\mathrm{DM}'$ has constructing element $\SGE{\Id}{-\tau}$ and therefore $\Z{4}$ charge 3.

\subsection{String interactions}

In order to find the three point couplings allowed by the space group selection rule~\cite{Hamidi:1986vh,Dixon:1986qv}, 
one needs to fulfill for each coupling that
\begin{align}
(\Id, 0) ~\in~[g_1]\cdot [g_2]\cdot [g_3]\;,
\end{align}
where $[g_i]$ denotes the conjugacy class of the constructing element $g_i$. The calculation is the 
same for the K\"ahler and the superpotential. In the K\"ahler potential, one looks for terms of the 
form $\Phi \Phi^\dagger V^{(\mathrm{M})}$, where $\Phi^\dagger$ has inverted quantum numbers and hence has the 
inverse constructing element. In the superpotential, one looks for terms of the form 
$\Phi_1 \Phi_2 \Phi_\mathrm{M}$, where $\Phi_2$ is either the mass partner of $\Phi_1$ (for the dark matter 
particle and the Higgs portal), or it is another field localized appropriately (for the neutrino portal).

In any case, we observe that there exist several winding states with trivial \Z4 charge, most 
prominently those with $n_2+n_4+n_6=0\Mod 2$. These states are particularly interesting candidates 
for mediators:
\begin{enumerate}
\item On the level of space group elements, they couple to both, DM and twisted strings. Let us 
work out for the coupling of dark matter to the $\theta$-twisted sector (i.e.\ $k=1$ and $\ell=0$ 
in eq.~\eqref{eq:bc}).
	\begin{itemize}
	\item It is evident that $\SGE{\Id}{ \frac{1}{2}(e_2-e_4-e_6)}\in \left[\SGE{\Id}{\tau}\right]$ and $\SGE{\Id}{ -\frac{1}{2}(e_2+e_4+e_6)}\in \left[\SGE{\Id}{-\tau}\right]$. Hence, $\SGE{\Id}{ 0}\in\left[\SGE{\Id}{\tau}\right]\cdot\left[\SGE{\Id}{-\tau}\right]\cdot\SGE{\Id}{ e_4+e_6}$.
	\item Similarly, $\SGE{\Id}{-\tau}\SGE{\theta}{0}\SGE{\Id}{\tau}=\SGE{\theta}{-e_4-e_6}\in\left[\SGE{\theta}{0}\right]$. Hence, $\SGE{\Id}{ 0}\in\SGE{\Id}{ e_4+e_6}\cdot\left[\SGE{\theta}{0}\right]\cdot\left[\SGE{\theta}{0}\right]$.
	\end{itemize}
\item Their local shift is a lattice vector, cf.\ ref.~\cite{Blaszczyk:2009in}. Hence, these states 
have $p_\mathrm{sh}=0$ and the corresponding couplings are not forbidden by gauge invariance.
\end{enumerate}

It turns out that the construction shown above not only works for the $\theta$-, but also for the 
$\omega$- and $\theta\omega$-twisted sector. In summary, we have the winding strings 
$V^{(\mathrm{M})}$ and $\Phi_\mathrm{M}$ that mediate between the dark matter strings ($\Phi_\mathrm{DM}$ 
and $\Phi_\mathrm{DM}'$) and the twisted sector 
\begin{table}[h]
\centering
\begin{tabular}{c||c|c|c}
sector of SM         & $\theta$&$\omega$&$\theta\omega$\\\hline
$g\in S$ of mediator & $\SGE{\Id}{ e_4+e_6}$&$\SGE{\Id}{ e_2+e_6}$&$\SGE{\Id}{e_2+e_4}$
\end{tabular}
\end{table}

In the next section, we will discuss the winding strings with constructing elements 
$\SGE{\Id}{\tau}$ and $\SGE{\Id}{ e_4+e_6}$ in more detail.

\subsection[Massive U(1) gauge bosons from string theory]{\boldmath Massive \U1 gauge bosons from string theory}
\label{sec:NarainDiscussion}

After the general discussion on the string origin of our dark matter candidate $\Phi_\mathrm{DM}$ 
and of the massive mediators $V^{(\mathrm{M})}$ and $\Phi_\mathrm{M}$, we now give more details on 
their existence and mass.

In heterotic string theory, a general string state is built out of independent right- and 
left-movers
\begin{equation}\label{eq:State}
\ket{p_\mathrm{R}; q}_\mathrm{R}\otimes\ket{p_\mathrm{L}; p_\mathrm{sh}}_\mathrm{L}\;,
\end{equation}
possibly subject to string oscillator excitations. Furthermore, $q$ is the bosonized right-moving 
$H$-momentum, being
\begin{equation}\label{eq:VM}
q~\in~\big\{ \big(\underline{\pm 1,0,0,0}\big) \;,\; \big(\pm\nicefrac{1}{2},\pm\nicefrac{1}{2},\pm\nicefrac{1}{2},\pm\nicefrac{1}{2}\big) \big\}\;.
\end{equation}
Here, the underline denotes all permutations and the number of plus-signs must be even for 
half-integer entries. In other words, $q$ is either an $\rep{8}_\mathrm{v}$ or an 
$\rep{8}_\mathrm{s}$ weight vector of \SO8 that fulfills $q^2=1$. Its first entry reflects the 
four-dimensional space-time chirality of the corresponding string state. In addition, the shifted 
left-moving momentum $p_\mathrm{sh}=p+A\,n$ in eq.~\eqref{eq:State} is given by the so-called 
discrete Wilson lines $A$~\cite{Ibanez:1986tp} and the momentum $p$ that belongs to the 
sixteen-dimensional $\E{8}\times\E{8}$ root lattice. Most important for our discussion are the 
right- and left-moving momenta, which are given by
\begin{subequations}\label{eq:pRpL}
\begin{eqnarray}
p_\mathrm{R} &:= &\frac{e^{-\mathrm{T}}}{\sqrt{2}} \left( \left(G-B+\frac{1}{2}A^\mathrm{T}A\right)\,n - m + A^\mathrm{T}p\right)\;,\label{eq:pR}\\
p_\mathrm{L} &:= &\frac{e^{-\mathrm{T}}}{\sqrt{2}} \left( \left(G+B-\frac{1}{2}A^\mathrm{T}A\right)\,n + m - A^\mathrm{T}p\right)\;,\label{eq:pL}
\end{eqnarray}
\end{subequations}
using the convention $\alpha'=1$, cf.\ ref.~\cite{GrootNibbelink:2017usl}. Here, similar to the 
discussion at the beginning of section~\ref{sec:StringRealization}, $e$ denotes the geometrical 
vielbein that defines the $D$-dimensional torus $\mathbb{T}^D$ with metric $G:=e^{\mathrm{T}}e$ and 
$B$ is the anti-symmetric $B$-field. Moreover, $n \in\Z{}^D$ are the integer winding numbers 
defined in analogy to eq.~\eqref{eq:bc} by the boundary condition of a bulk string
\begin{equation}\label{eq:bc2}
X(\sigma_0, \sigma_1+1) ~=~ X(\sigma_0,\sigma_1) + e\,n\;, \quad\mathrm{where}\quad e\,n ~=~ \frac{1}{\sqrt{2}}\left(p_\mathrm{R}+p_\mathrm{L}\right)\;,
\end{equation}
and $m\in\Z{}^D$ denote the integer Kaluza--Klein (KK) numbers. Note that the $(2D+16)$-dimensional 
vectors $(p_\mathrm{R},p_\mathrm{L},p_\mathrm{sh})$ span an even, integer and self-dual lattice 
with signature $(D,D+16)$, called the Narain lattice. As such, a vector given by eqs.~\eqref{eq:pRpL} 
satisfies for example
\begin{equation}\label{eq:NarainSP}
-(p_\mathrm{R})^2 + (p_\mathrm{L})^2 + (p_\mathrm{sh})^2~=~ 2\,m^\mathrm{T} n + p^2~=~ \mathrm{even}\;,
\end{equation}
reflecting the fact that the Narain lattice is even.

A physical string state from eq.~\eqref{eq:State} is subject to the so-called level-matching 
condition on right- and left-moving masses, i.e.
\begin{equation}\label{eq:LM}
M_\mathrm{R}^2 ~=~ M_\mathrm{L}^2\;,
\end{equation}
where
\begin{equation}\label{eq:RLMMass}
\frac{1}{2} M_\mathrm{R}^2 ~=~ (p_\mathrm{R})^2 + q^2 + 2 \left(N_\mathrm{R} - \frac{1}{2}\right) \quad,\quad \frac{1}{2} M_\mathrm{L}^2 ~=~ (p_\mathrm{L})^2 + (p_\mathrm{sh})^2 + 2 \left(N_\mathrm{L} - 1\right)\;. 
\end{equation}

We are interested in winding strings in order to discuss the origin of both, our dark matter 
candidate $\Phi_\mathrm{DM}$ with constructing element $\SGE{\Id}{\tau}\in S$ and the mediators, 
for example, $V^{(\mathrm{M})}$ and $\Phi_\mathrm{M}$ with constructing element 
$\SGE{\Id}{ e_4+e_6}\in S$. To do so, we can concentrate on three compactified dimensions $D=3$ and 
focus on the torus directions $e_2$, $e_4$ and $\tau$, see eq.~\eqref{eq:tau}. To keep the 
discussion short we assume trivial Wilson lines $A_2=A_4=A_\tau=(0^{16})$. Then, we can consider 
the $\Z{2}\times\Z{2}$ orbifold of this $\mathbb{T}^3$ subsector in order to analyze those winding 
strings we are mostly interested in.

For the $D=3$ subsector of the $\Z2\times\Z2$-5-1 orbifold geometry, the Narain lattice 
eq.~\eqref{eq:pRpL} can be parameterized by three radii $R_2$, $R_4$ and $R_6$ for the torus 
vielbein $e$ and three parameters $b_1$, $b_2$ and $b_3$ for the anti-symmetric $B$-field, i.e.\
\begin{equation}\label{eq:ModuliParameters}
e ~=~ \left(\begin{array}{ccc}R_2 & 0 & \nicefrac{R_2}{2}\\ 0 & R_4 & \nicefrac{R_4}{2}\\ 0 & 0 & \nicefrac{R_6}{2}\end{array}\right) \quad,\quad B ~=~ \left(\begin{array}{ccc} 0 & b_1 & b_2\\-b_1 & 0 & b_3\\-b_2 & -b_3 & 0\end{array}\right)\;.
\end{equation}
Thus, the columns of the geometrical vielbein $e$ are given by $e_2$, $e_4$ and $\tau$, cf.\ 
eq.~\eqref{eq:tau}. Consequently, eq.~\eqref{eq:ModuliParameters} has six free parameters (which 
combine in the six-dimensional $\Z{2}\times\Z{2}$ orbifold with six additional parameters to three 
K\"ahler moduli $T_i$ and three complex structure moduli $U_i$, $i=1,2,3$).

Let us begin with the discussion on the mediator with constructing element $\SGE{\Id}{ e_4+e_6}$. 
In terms of the basis $e_2$, $e_4$ and $\tau$, we use $e_6 = 2\tau-e_2-e_4$ to write 
$\SGE{\Id}{ e_4+e_6} = \SGE{\Id}{ 2\tau-e_2}$. Hence, the mediator has winding numbers 
$n=(-1,0,2)^\mathrm{T}$ such that $e\,n=2\tau-e_2$. It turns out that there exists a point in 
moduli space (i.e.\ with special values for the radii $R_i$ and $B$-field components $b_i$), where 
the mediator $\SGE{\Id}{2\tau-e_2}$ becomes massless. Thus, we start our discussion at this point in 
moduli space and, afterwards, move in moduli space to make the mediator massive. 

Massless strings must have vanishing right- and left-moving masses eqs.~\eqref{eq:RLMMass}, subject 
to $M_\mathrm{R}^2 = M_\mathrm{L}^2$. A vanishing right-moving mass implies $p_\mathrm{R} = (0^3)$ 
and, hence, 
\begin{equation}\label{eq:pRZeroCondition}
m ~=~ (G-B)\,n\;, \quad\mathrm{using}\quad A_2~=~A_4~=~A_\tau=(0^{16})
\end{equation}
and $q^2 = 1$ and $N_\mathrm{R}=0$ in eq.~\eqref{eq:RLMMass}. In this case, $p_\mathrm{L}$ is 
given by $p_\mathrm{L} = \sqrt{2}\,e\,n$. Consequently, a vanishing left-moving mass, 
eq.~\eqref{eq:RLMMass} with $p=(0^{16})$ and $N_\mathrm{L}=0$, yields the constraint
\begin{equation}\label{eq:R4R6MassCondition}
n^\mathrm{T} G\,n ~=~ 1 \quad\Leftrightarrow\quad (R_4)^2 + (R_6)^2 ~=~ 1\;,
\end{equation}
for our mediator string with winding numbers $n=(-1,0,2)^\mathrm{T}$. Note that one can check that 
this mass condition is identical for all four winding strings in the conjugacy class 
$[\SGE{\Id}{ 2\tau-e_2}]$, as expected. In order to satisfy the mass 
condition~\eqref{eq:R4R6MassCondition}, we choose
\begin{equation}\label{eq:ModuliSpace1}
(R_4)^2 ~=~ (R_6)^2 ~=~ \nicefrac{1}{2}\;.
\end{equation}
Next, we have to fix the remaining moduli parameters $b_i$ such that the KK numbers $m$ in 
eq.~\eqref{eq:pRZeroCondition} become integer. In other words, a general winding string 
$n\neq (0^3)$ necessarily carries non-trivial KK numbers $m\neq (0^3)$ in order to satisfy 
eq.~\eqref{eq:pRZeroCondition} and, hence, level-matching. We find a solution for
\begin{equation}\label{eq:ModuliSpace2}
b_1 ~=~ 0\quad,\quad b_2 ~=~ \nicefrac{1}{2}\quad\mathrm{and}\quad b_3 = -\nicefrac{1}{4}\;,
\end{equation}
such that $m = (-1,1,0)^\mathrm{T}$ satisfies eq.~\eqref{eq:pRZeroCondition}. Let us give two 
import remarks: First, at a generic point in moduli space the total mass squared 
$M_\mathrm{R}^2 + M_\mathrm{L}^2$ of the mediator depends on all six free parameters $b_i$, 
$i=1,2,3$, $R_2$, $R_4$ and $R_6$. However, there are special points in moduli space, where the 
mass is independent of, for example, the compactification radius $R_2$: this is the case at 
$b=\nicefrac{1}{2}$. Secondly, since $m^\mathrm{T} n = 1$ and $p=(0^{16})$, the even Narain lattice 
ensures that $-(p_\mathrm{R})^2+(p_\mathrm{L})^2 = 2$, see eq.~\eqref{eq:NarainSP}. Hence, the 
level-matching condition~\eqref{eq:LM} is satisfied everywhere in moduli space for this string.

\begin{figure}[t!]
\centering
\includegraphics[width=.5\textwidth]{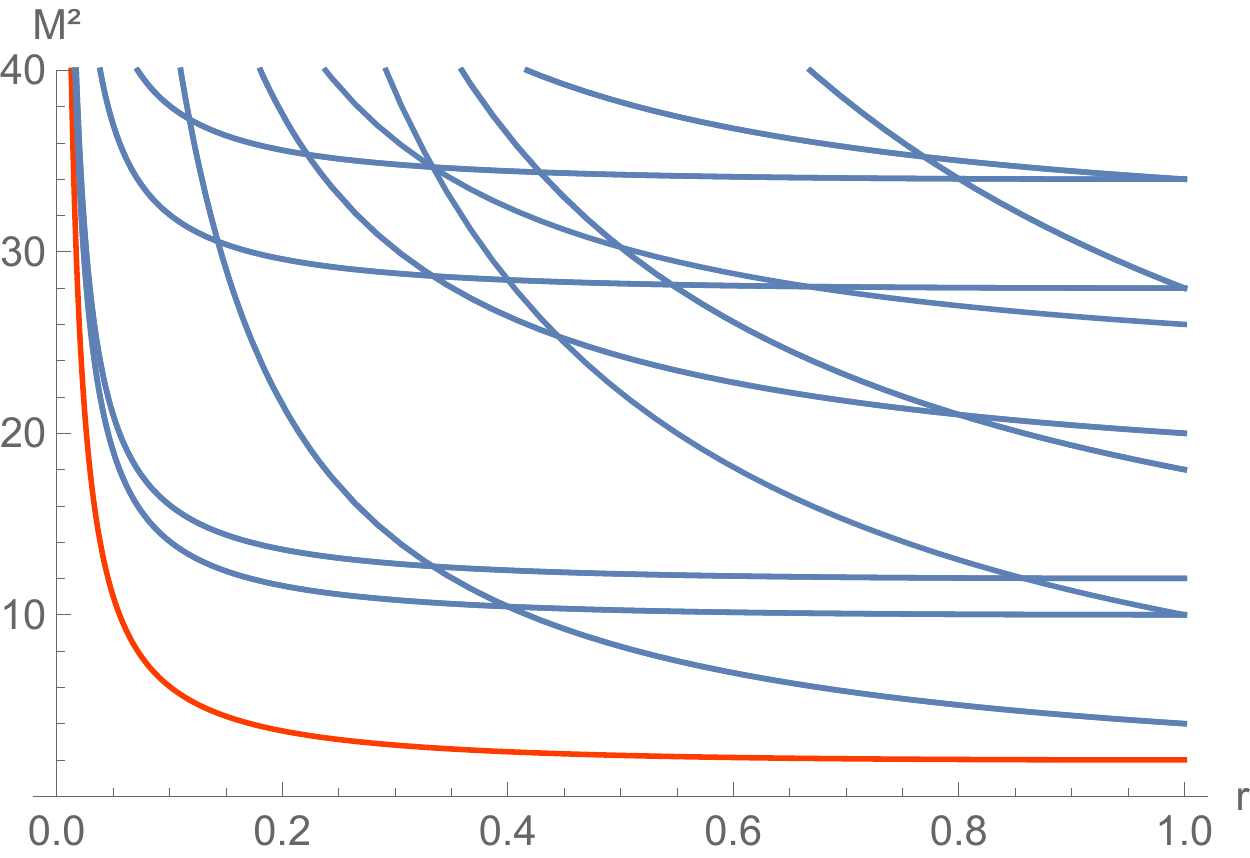}
\vspace{-0.2cm}
\caption{The mass squared $M^2$ (in string units $\alpha'=1$) of dark matter candidates with 
constructing element $\SGE{\Id}{\tau}\in S$ and various KK numbers $-5 \leq m_i \leq 5$ for $i=1,2$ 
(and $m_3=1$) depending on the compactification radius $r=(R_2)^2$. The lightest $\tau$-winding 
string state (in red) is specified by $n=(0,0,1)^\mathrm{T}$ and $m=(0,1,1)^\mathrm{T}$. In this 
figure, the other moduli parameters are set according to eqs.~\eqref{eq:ModuliSpace1} 
and~\eqref{eq:ModuliSpace2}, where the mediator is massless independently of $r=(R_2)^2$. There 
are other points in moduli space, where the masses of the lightest dark matter candidate and of the 
mediator are much smaller than 1.}
\label{Fig:Tau}
\end{figure}

Next, we consider the dark matter candidate $\Phi_\mathrm{DM}$ with constructing element 
$\SGE{\Id}{\tau}\in S$. In this case, the corresponding winding numbers are given by 
$n=(0,0,1)^\mathrm{T}$ such that $e\,n=\tau$. Since we are interested in the lightest string state 
with these winding numbers, we set $N_\mathrm{R} = N_\mathrm{L} = 0$ and $p=(0^{16})$. Note that 
$p$ defines the representation of $\Phi_\mathrm{DM}$ under the four-dimensional gauge group, which 
originates from $\E{8}\times\E{8}$ and is assumed to contain the Standard Model gauge group. Hence, 
$p=(0^{16})$ renders $\Phi_\mathrm{DM}$ a Standard Model singlet. Now, we have to find KK numbers 
$m\in\Z{}^3$ such that $m^\mathrm{T} n = 1$. Then, the Narain condition 
$-(p_\mathrm{R})^2+(p_\mathrm{L})^2 = 2m^\mathrm{T} n = 2$ ensures level-matching. Hence, 
$m = (m_1, m_2,1)^\mathrm{T}$ with $m_1$, $m_2\in\Z{}$ is the general solution. Now, let us find the 
lightest dark matter candidate $\Phi_\mathrm{DM}$. To do so, we assume eqs.~\eqref{eq:ModuliSpace1} 
and~\eqref{eq:ModuliSpace2}, and compute the total mass squared of $\Phi_\mathrm{DM}$ in terms of 
the free radius $r=(R_2)^2$ and KK numbers $m_1$, $m_2\in\Z{}$,
\begin{eqnarray}
M^2(r,m_1,m_2)  & \propto & (p_\mathrm{R})^2+(p_\mathrm{L})^2 - 2\\
                & \propto & 18+8m_1^2+16m_2(m_2-2)+4m_1(4m_2-7)+ \frac{(1+2m_1)^2}{r}+r\;.\nonumber
\end{eqnarray}
Let us constrain the radius $R_2$ to $0 < R_2 < 1$. In this range, the $\tau$-winding string with 
minimal mass has KK numbers $m=(0,1,1)^\mathrm{T}$: in figure~\ref{Fig:Tau} we plot the masses as 
functions of $r=(R_2)^2$ for various $\tau$-winding strings with KK numbers in the ranges 
$-5 \leq m_i \leq 5$ for $i=1,2$ and identify the lightest string.

Consequently, we have identified the lightest $\tau$-winding string state, specified by winding and 
KK numbers
\begin{equation}
n~=~ (0,0,1)^\mathrm{T} \quad\mathrm{and}\quad m ~=~ (0,1,1)^\mathrm{T}\;,
\end{equation}
respectively. This massive string is our dark matter candidate $\Phi_\mathrm{DM}$. It is a Standard 
Model singlet, i.e.\ $p=(0^{16})$, and stable since it carries an odd $\Z{4}$ charge $Q=1$, cf. 
eq.~\eqref{eq:Z4Charge}. Furthermore, the lightest mediator corresponding to a winding string with 
constructing element $\SGE{\Id}{ e_4+e_6} \in S$ is characterized by winding and KK numbers
\begin{equation}
n~=~ (-1,0,2)^\mathrm{T} \quad\mathrm{and}\quad m ~=~ (-1,1,0)^\mathrm{T}\;,
\end{equation}
respectively. In contrast to the the dark matter candidate, the mediator is uncharged under $\Z{4}$, 
cf. eq.~\eqref{eq:Z4Charge}. As we have shown, if we keep $b_2=\nicefrac{1}{2}$ fixed, we can 
independently vary the masses of $\Phi_\mathrm{DM}$ and the mediator. Finally, it is important to 
comment that the mediator corresponds to a $\U{1}$ gauge boson that becomes massless at a specific 
point in moduli space, given in eqs.~\eqref{eq:ModuliSpace1} and~\eqref{eq:ModuliSpace2}. Moreover, 
using the results of ref.~\cite{Beye:2014nxa}, we know that the massive $\tau$-winding string state 
is charged under this $\U{1}$. Hence, a K\"ahler potential of the form eq.~\eqref{eq:KahlerPot} 
must originate from this string construction.

\providecommand{\href}[2]{#2}\begingroup\raggedright\endgroup\end{document}